\documentclass[twocolumn,preprintnumbers,amsmath,amssymb]{revtex4}

\usepackage{graphicx,color}%

\begin{document}

\title{Scale and conformal invariance in field theory: a physical counterexample}

\author{V. Riva$^{a,b}$ and J. Cardy$^{a,c}$}

\affiliation{$^{a}$Rudolf Peierls Centre for Theoretical Physics,
1 Keble Road, Oxford OX1 3NP, UK} \affiliation{$^{b}$Wolfson
College, Oxford} \affiliation{ $^{c}$All Souls College, Oxford}

\begin{abstract}
\par
In this note, we illustrate how the two--dimensional theory of
elasticity provides a physical example of field theory displaying
scale but not conformal invariance.
\end{abstract}


\maketitle


\section{Introduction}

In the quantum field theory literature, scale invariance is often
assumed to imply conformal invariance, provided the theory is
local. Furthermore, both invariances are usually considered
equivalent to the tracelessness of the stress--energy tensor.
These widely held convictions, sustained by the difficulty of
finding counterexamples, are actually incorrect.

Coleman and Jackiw \cite{jackiw} clarified this issue in the case
of four space--time dimensions, showing that conformal invariance
is not in general guaranteed by the presence of scale invariance.
A systematic analysis of the problem for arbitrary dimensionality
$D$ was then performed by Polchinski \cite{polchinski}, who
achieved the same conclusion for any $D\neq 2$. In the particular
case $D=2$, however, Polchinski proved that scale invariance
implies conformal invariance under broad conditions. In the
following, we will focus on this interesting dimensionality,
providing a physical example in which the implication does not
hold.

Let us now summarize the observations presented in
Ref.\,\cite{polchinski}. Given a symmetric and conserved
stress--energy tensor $T_{\mu\nu}(x)$, the property of scale
invariance can be equivalently formulated in terms of its trace as
\begin{equation}\label{scaleinv}
T_{\mu}^{\;\;\mu}(x)=-\partial_{\mu}K^{\mu}(x)\;,
\end{equation}
where $K^{\mu}(x)$ is some local operator. Conformal invariance
further requires the existence of another local operator $L(x)$
such that
\begin{equation}\label{confinv}
K_{\mu}(x)=-\,\partial_{\mu}L(x) \quad \Rightarrow\quad
T_{\mu}^{\;\;\mu}(x)=\partial_{\mu}\partial^{\mu} L(x)\;.
\end{equation}
The above property is then equivalent to the tracelessness of the
stress--energy tensor, because one can define the \lq improved'
tensor
\begin{equation}\label{improved}
\Theta_{\mu\nu}(x)\,=\,T_{\mu\nu}(x)\,+\,\partial_{\mu}\partial_{\nu}L(x)\,-\,g_{\mu\nu}\,
\partial_{\rho}\partial^{\rho}L(x)\;,
\end{equation}
which is both conserved and traceless. As properly emphasized in
Ref.\,\cite{polchinski}, most of the physically relevant theories
display both scale and conformal invariance because they do not
have any non--trivial candidate for $K_{\mu}$. We will see in the
following how this is the crucial ingredient in our
counterexample.

Besides these general remarks, Polchinski also refined an argument
by Zamolodchikov \cite{cth}, demonstrating that scale invariance
implies conformal invariance in $D=2$. The proof consists of
defining another kind of \lq improved' stress--energy tensor
$\Theta^{'}_{\mu\nu}(x)$, whose trace is shown to have a vanishing
two--point function:
\begin{equation}\label{2pttrace}
\langle
\Theta_{\mu}^{'\;\;\mu}(x)\,\Theta_{\sigma}^{'\;\;\sigma}(0)\rangle\,=\,0\;.
\end{equation}
The sufficient condition for constructing $\Theta^{'}_{\mu\nu}(x)$
is a discrete spectrum of scaling dimensions, and, together with
the assumption of reflection positivity, (\ref{2pttrace}) implies
the vanishing of the trace $\Theta^{'\;\;\mu}_{\mu}$ itself.
Actually, under the above hypotheses the two \lq improved' tensors
$\Theta_{\mu\nu}(x)$ and $\Theta^{'}_{\mu\nu}(x)$ coincide.

\section{The model}

Let us now introduce a physical example in which scale invariance
does not imply conformal invariance. This is the theory of
elasticity \cite{landau} in two dimensions, defined by the
Euclidean action
\begin{equation}\label{elasticity}
{\cal S}\,=\,\int\,d^{2}x\;{\cal
L}\,=\,\frac{1}{2}\,\int\,d^{2}x\,\left\{2\,g\,
u_{\mu\nu}\,u^{\mu\nu}+k\, (u_{\sigma}^{\;\;\sigma})^2\right\}\;,
\end{equation}
where $\;u_{\mu\nu}=\frac{1}{2}\,\left(\partial_{\mu}
u_{\nu}+\partial_{\nu} u_{\mu}\right)\;$ is the so--called strain
tensor, built with the \lq displacement fields' $u_{\mu}$. Greek
indices run over $1,2$ and we use the summation convention. The
coefficients $g$ and $k+g$ represent, respectively, the shear
modulus and the bulk modulus of the described material.

The action (\ref{elasticity}) is invariant under translations,
rotations and dilatations, provided the fields $u_{\mu}$ transform
under rotations
$\,x'^{\;\mu}\,=\,\Lambda^{\mu}_{\;\;\nu}\,x^{\nu}\,$ as vectors
\begin{equation}\label{rotation}
u'_{\mu}(x')\,=\,\Lambda_{\mu}^{\;\;\nu}\,u_{\nu}(x)\;,
\end{equation}
while no change is required for fields under dilatations. The
canonical stress--energy tensor
\begin{equation}\label{canonical}
T^{\,\text{c}}_{\mu\nu}\,=\,\frac{\partial{\cal
L}}{\partial(\partial^{\mu}u_{\sigma})}\,\partial_{\nu}u_{\sigma}\,-\,g_{\mu\nu}\,{\cal
L}
\end{equation}
associated to (\ref{elasticity}) is traceless but not symmetric.
However, a symmetric and conserved tensor $T_{\mu\nu}$ can be
conventionally constructed via the Belinfante prescription:
\begin{equation}\label{Belinfante}
T_{\mu\nu}\,=\,T^{\,\text{c}}_{\mu\nu}\,+\,\partial^{\rho}B_{\rho\mu\nu}\;,
\end{equation}
where
\begin{widetext}
\begin{equation}\label{defB}
B_{\rho\mu\nu}\,=\,\frac{i}{2}\,\left\{\frac{\partial{\cal
L}}{\partial(\partial^{\rho}u_{\sigma})}\,S_{\nu\mu}\,u_{\sigma}+\frac{\partial{\cal
L}}{\partial(\partial^{\mu}u_{\sigma})}\,S_{\rho\nu}\,u_{\sigma}+\frac{\partial{\cal
L}}{\partial(\partial^{\nu}u_{\sigma})}\,S_{\rho\mu}\,u_{\sigma}\right\}\,=\,-\,B_{\mu\rho\nu}\;.
\end{equation}
\end{widetext}
$S_{\mu\nu}$ is an antisymmetric tensor, taking values in the
representations of the Lorentz group, which expresses the
variation of the field multiplet $\phi=\{u_{\mu}\}$ under
infinitesimal rotations $\;x'\,^{\mu}\,\simeq
\,x^{\mu}+\omega^{\mu}_{\;\;\nu}\,x^{\nu}\;$:
\begin{equation}\nonumber
\phi'(x')\,\simeq\,\left(I-\frac{i}{2}\,\omega_{\rho\nu}S^{\rho\nu}\right)\phi(x)\;.
\end{equation}
In our case the fields transform according to the vector
representation (\ref{rotation}), and the only non--vanishing
Euclidean components of $S_{\mu\nu}$ act as
\begin{eqnarray*}
S_{12}\,u_1\,=\,-\,S_{21}\,u_1&=&i\,u_2\\
S_{12}\,u_2\,=\,-\,S_{21}\,u_2&=&-\,i\,u_1
\end{eqnarray*}

It follows from (\ref{Belinfante}) that the trace of the
stress--energy tensor can be cast in the form (\ref{scaleinv})
\begin{equation}\label{noconf}
T_{\mu}^{\;\;\mu}\,=\,-\partial^{\mu}K_{\mu}\qquad\quad\text{with}\qquad\quad
K_{\mu}\,=\,-B_{\mu\rho}^{\;\;\;\;\rho}\;,
\end{equation}
in agreement with the scale invariance of the theory. In order to
investigate whether the additional property (\ref{confinv}),
equivalent to conformal invariance, is also attained, it is now
convenient to explicitly write $K_{\mu}$ in Euclidean coordinates.
We have
$$
K_1=\partial_1\left[-\,\frac{k}{2}\,u_1^2+\frac{g}{2}\,u_2^2\right]
-(k+2g)\,u_1\partial_2 u_2+g \,u_2\partial_2 u_1 \nonumber
$$\vspace{-0.5cm}
\begin{equation}\label{Beucl}
K_2=\partial_2\left[\frac{g}{2}\,u_1^2-\frac{k}{2}\,u_2^2\right]
-(k+2g)\,u_2\partial_1 u_1+g \,u_1\partial_1 u_2 \;.
\end{equation}
It appears from (\ref{Beucl}) that $K_{\mu}$ cannot be entirely
reduced to a gradient\footnote{As clearly explained in
\cite{polchinski}, no further freedom in the definition of the
stress--energy tensor can modify this conclusion.}, therefore the
necessary condition (\ref{confinv}) for conformal invariance does
not hold and the stress--energy tensor cannot be \lq improved' to
be traceless.

\vspace{0.5cm}

The lack of conformal invariance in (\ref{elasticity}) becomes
manifest if we write the action in complex coordinates $\;z=x^1+i
x^2\;,\;\bar{z}=x^1-i x^2\;$. We have

\begin{equation}\label{elasticitycomplex}
{\cal
S}\,=\,\frac{1}{2}\,\int\,d^{2}z\,\left\{(k+g)\,\left[\partial
\bar{u}+\bar{\partial} u\right]^2+4\,g\,(\partial
u)(\bar{\partial} \bar{u})\right\}\;,
\end{equation}
where $\;\partial=\frac{\partial}{\partial
z}\;,\;\bar{\partial}=\frac{\partial}{\partial \bar{z}}\;$ and
$\;u=u_1-i u_2\;,\;\bar{u}=u_1+i u_2\;$. The transformation
(\ref{rotation}) under rotations translates into the requirement
that the fields $u$ and $\bar{u}$ have spins $s_u=1$ and
$s_{\bar{u}}=-1$, while both their scaling dimensions
$\Delta_{u}\,,\,\Delta_{\bar{u}}\;$ have to vanish in order to
ensure scale invariance. These properties are obtained by
assigning the conformal weights
\begin{equation}\label{confweights}
h_{u}=\bar{h}_{\bar{u}}=\frac{1}{2}\;,\qquad
\bar{h}_{u}=h_{\bar{u}}=-\,\frac{1}{2}\;,
\end{equation}
which are defined through
$$
\Delta=h+\bar{h}\;,\qquad \qquad s=h-\bar{h}\;.
$$
It is then easy to see that (\ref{elasticitycomplex}) is not
invariant under a conformal transformation $\;z\to
w=f(z)\;,\;\bar{z}\to \bar{w}=\bar{f}(\bar{z})\;$, where the
fields transform as $\;\phi \to
(f')^{-h}\,(\bar{f}')^{-\bar{h}}\;\phi\;$.

Conformal invariance is only recovered in the unphysical case of
zero bulk modulus $k+g=0$, when (\ref{elasticitycomplex})
describes a conformal field theory with central charge $c=2$. This
is the familiar situation in which the two fields $u$ and
$\bar{u}$ are not required to transform under rotations, and the
corresponding symmetry is then enlarged from $O(2)$ to $O(2)\times
O(2)$. It is worth stressing that all conformal weights
(\ref{confweights}) now vanish, therefore this particular case
cannot be described by simply evaluating the previous results in
the limit $k+g\to 0$, but must be separately treated. In fact, the
canonical stress--energy tensor (\ref{canonical}) is now already
symmetric, therefore no Belinfante construction has to be
implemented, and conformal invariance immediately follows from the
tracelessness of $T^{\,\text{c}}_{\mu\nu}$.

\vspace{0.5cm}

As a final remark, we will now show that the above observations
for generic values of $k+g$ are not in contradiction with the
statement, proven in \cite{polchinski}, that the stress--energy
tensor can be nevertheless \lq improved' to a certain degree, in
order to obtain a vanishing two--point function of its trace, as
in (\ref{2pttrace}). However, this is not associated with the
vanishing of the trace itself, because the theory under
examination turns out to be not reflection positive, as we shall
illustrate below.

Let us first notice that the trace of $T_{\mu\nu}$ can be
expressed at quantum level as
\begin{eqnarray}\nonumber
T_{\mu}^{\;\;\mu}&=&(k+g)\,\left[\,:\partial\bar{u}\,\partial\bar{u}:\,+\,
:\bar{\partial}u\,\bar{\partial}u:\,+\,
2\,:\partial\bar{u}\,\bar{\partial}u:\,\right]\,+\,\\
&&\,-\,g\,\left[\,:\partial u \,\bar{\partial}\bar{u}:\,-\, : u\,
\partial\bar{\partial}\bar{u}:\,-\,: \bar{u}
\,\partial\bar{\partial}u:\,\right]\;,\label{tracebel}
\end{eqnarray}
where the symbol \lq $:\;:$' indicates normal ordering. By using
Wick theorem and the explicit expressions
\begin{eqnarray}\label{2ptcomplex}
\langle \,u(z)\,u(w)\,
\rangle&=&\frac{k+g}{4\pi\,g\,(k+2g)}\,\frac{\bar{z}-\bar{w}}{z-w}\;\;,\nonumber\\
\langle\, \bar{u}(z)\,\bar{u}(w)\,
\rangle&=&\frac{k+g}{4\pi\,g\,(k+2g)}\,\frac{z-w}{\bar{z}-\bar{w}}\;\;,\\
\langle\, u(z)\,\bar{u}(w)\,\rangle&=&\frac{k+g-(k+3g)\,\log
(z-w)(\bar{z}-\bar{w})}{4\pi\,g\,(k+2g)}\;\;,\nonumber
\end{eqnarray}
it is then straightforward to check that the two--point function
of (\ref{tracebel}) does not vanish, being
\begin{equation}
\langle T_{\mu}^{\;\;\mu}(z)
T_{\sigma}^{\;\;\sigma}(0)\rangle\,=\,
\frac{-32\,g^2(k+g)(k+3g)}{[4\pi g(k+2g)]^2}
\,\frac{1}{z^2\bar{z}^2}\;.
\end{equation}

However, we can guess the kind of improvement to be performed on
$T_{\mu\nu}$, by observing that the operator $K_{\mu}$ defined in
(\ref{noconf}) can be partially reduced to a gradient, as
\begin{eqnarray}\nonumber
K_{z}&=&\partial\,(g\,u\bar{u})-\frac{k+g}{2}\;u\,\bar{\partial}
u-\frac{k+3g}{2}\; u\,\partial
\bar{u}\;,\\
K_{\bar{z}}&=&\bar{\partial}\,(g\,u\bar{u})-\frac{k+g}{2}\;\bar{u}\,\partial
\bar{u}-\frac{k+3g}{2}\; \bar{u}\,\bar{\partial} u\;.
\end{eqnarray}
It is then natural to define $\Theta'_{\mu\nu}$ as in
(\ref{improved}), with $L\,=\,-g\,u\bar{u}$, and it can be easily
checked that the two--point function of its trace indeed vanishes
\begin{equation}\label{2pttheta}
\langle\Theta_{\mu}^{'\;\;\mu}(z)\Theta_{\sigma}^{'\;\;\sigma}(0)\rangle\,=\,0\;,
\end{equation}
although the trace itself does not
\begin{eqnarray}\nonumber
\Theta_{\mu}^{'\;\;\mu}&=&(k+g)\,\left[\,:\partial
\bar{u}\,\partial\bar{u}:\,+\, :\bar{\partial}
u\,\bar{\partial}u:\,\right]\,+\,\\
&& \,+\,2\,(k+3g)\,:\partial \bar{u}\,\bar{\partial}u:\;.
\end{eqnarray}
suggesting therefore the failure of reflection positivity. In
fact, in a reflection positive theory eq.\,(\ref{2pttheta}) should
imply the vanishing of any two--point function involving
$\Theta_{\mu}^{'\;\;\mu}$, but several counterexamples occur here,
for instance
$$
\langle \Theta_{\mu}^{'\;\;\mu}(z)\,:\partial u\partial
u:(0)\rangle\,=\,-\,\frac{k+g}{2\pi^2\,g\,(k+2g)}\,\frac{1}{z^4}
\;.
$$

The lack of reflection positivity can be equivalently seen as
non--unitarity in Minkowski coordinates. In fact, the Hamiltonian
associated to (\ref{elasticity}) is not positive definite, being
expressed as
\begin{eqnarray}\label{ham}
H&=&\frac{1}{2}\,\int\,dx\,\left\{\frac{1}{k+2g}\,\pi_t^2\,+\,g\,(\partial_x
u_t)^2\,+\,\right.\\
&&\left.\,-\,\frac{1}{g}\,[\pi_x-(k+g)\,\partial_x
u_t]^2\,-\,(k+2g)(\partial_x u_x)^2\right\}\;,\nonumber
\end{eqnarray}
where the conjugate momenta to $u_t,u_x$ are given by
\begin{eqnarray}\nonumber
\pi_{t}&=&(k+2g)\partial_{t} u_{t}\;,\\
\pi_x&=&g\,\partial_{t} u_x+(k+g)\,
\partial_x u_{t}\;.\nonumber
\end{eqnarray}
The negative signs in (\ref{ham}) are produced by the $(1,-1)$
signature of the Minkowski target space $\{u_t,u_x\}$, which
follows from the transformation property (\ref{rotation}).

\section{Comments}

In concluding this note, it is worth emphasizing that the lack of
conformal invariance in the discussed example entirely originates
from the transformation property (\ref{rotation}) of the fields
under rotations. This is what makes the canonical stress--energy
tensor $T^{\,\text{c}}_{\mu\nu}$ not symmetric and provides the
non--trivial expression (\ref{noconf}) for the trace of the
symmetrized tensor $T_{\mu\nu}$. The symmetrization procedure
always respects scale invariance, since the Belinfante
prescription only contributes with the divergence of a local
function $K_{\mu}$ to the trace $T_{\mu}^{\;\;\mu}$, as shown in
(\ref{noconf}). However, $K_{\mu}$ has in general a non--trivial
vectorial structure, and therefore $T_{\mu\nu}$ cannot be made
traceless, consistently with the fact that the theory is not
conformally invariant.

Finally, it is interesting also to notice that the non--unitarity
of the theory, which reconciles our observations with Polchinski's
proof, is itself a consequence of the vectorial nature of the
fields $u_{\mu}$. In fact, covariance requires both worldsheet and
target space to have the same Lorentzian signature, making the
Hamiltonian (\ref{ham}) not positive definite. A similar
phenomenon takes place also in electrodynamics and string theory,
where the Lorentzian  signature of target space produces
negative--norm states in the Hilbert space. However, in those
cases the presence of a gauge invariance allows one to recover
unitarity in the subspace of physical states.

\vspace{0.5cm}

\textbf{Acknowledgments}

We thank R. Jackiw for bringing relevant literature to our
attention. V.R. is also grateful to L. Brualla for useful
comments. This work was supported by EPSRC, under the grant
GR/R83712/01.

\end{document}